# Beyond the Hype: A Cautionary Tale of ChatGPT in the Programming Classroom

Grant Oosterwyk[0000-0002-2745-3929],Pitso Tsibolane[0000-0002-7888-1181], Popyeni Kautondokwa[0000-0001-9001-7313], and Ammar Canani

Department of Information Systems, Faculty of Commerce, University of Cape Town, Cape Town, South Africa

grant.oosterwyk@uct.ac.za, pitso.tsibolane@uct.ac.za, ktnpop001@myuct.ac.za & cnnamm001@myuct.ac.za

**Abstract.** Due to the proliferation of Large Language Models research and the use of various Artificial Intelligence (AI) tools, the field of information systems (IS) and computer science (CS) has evolved. The use of tools such as ChatGPT to complete various student programming exercises (e.g., in Python) and assignments has gained prominence amongst various academic institutions. However, recent literature has suggested that the use of ChatGPT in academia is problematic and the impact on teaching and learning should be further scrutinized. More specifically, little is known about how ChatGPT can be practically used with code (programming) writing to complete programming exercises amongst IS and CS undergraduate university students. Furthermore, the paper provides insights for academics who teach programming to create more challenging exercises and how to engage responsibly in the use of ChatGPT to promote classroom integrity. In this paper, we used Complex Adaptive Systems (CAS) theory as a theoretical guide to understand the various dynamics through classroom code demonstrations. Using ChatGPT 3.5, we analyzed the various practical programming examples from past IS exercises and compared those with memos created by tutors and lecturers in a university setting. This paper highlights common ways of assessment, programming errors created by ChatGPT and the potential consideration for IS academics to ensure the development of critical programming skills among students.

**Keywords:** ChatGPT, Large Language Models, Artificial Intelligence, Complex Adaptive Systems.

## 1 Introduction

With the proliferation of Large Language Models (LLM) and their use in research and practice, many organisations are starting to leverage artificial intelligence (AI) tools such as ChatGPT for competitive advantage (Dwivedi et al. 2023). In various cases, the value of these tools gives way to enhancing productivity among practitioners. However, in the context of academia, the use of AI tools for educational purposes, specifically related to problem-solving and critical thinking (Matthee and Turpin, 2019) in software programming, has become increasingly problematic in the information systems (IS) and computer science (CS) disciplines.

LLMs, such as ChatGPT, can be defined as a human-text generator capable of scanning text patterns of large internet datasets using Natural Language Processing (NLP) (Rahman et al. 2023). A recent editorial paper has highlighted that ChatGPT is not limited to natural language (Dwivedi et al. 2023) and that the tool is able to handle various querying languages (e.g., Python, Java, C++ and others). The paper further recommended that tasks should be designed to be more challenging given the current design of many teaching pedagogies. Other studies (Shahrasbi, Jin, and Zheng, 2021) critically analyse traditional pedagogical methodologies employed in programming and application development courses, which tend to emphasize programming exercises while neglecting opportunities for meaningful student interaction. As a countermeasure, the researchers put forth an innovative instructional framework for a mobile application development course, predicated upon a design thinking paradigm that capitalizes on the active involvement of students within the design process.

Hence, one of the objectives of this paper is to demonstrate how ChatGPT can be used to write basic programming tasks to complete various university programming assignments (i.e., exercises). The paper further aims to investigate the challenges associated with using ChatGPT for completing programming tasks using an IS foundation course and argues for the need to design more challenging assignments that foster critical programming skills among students. Furthermore, we employed the Complex Adaptive Systems (CAS) theory as a guiding framework to comprehend the multifaceted dynamics present in demonstrations.

Using ChatGPT, we scrutinized various practical programming examples derived from past programming exercises and juxtaposed them with memos crafted by tutors and lecturers at a higher education institution. This paper elucidates prevalent assessment methods, programming inaccuracies generated by ChatGPT and potential considerations for IS educators to facilitate the cultivation of critical programming competencies in their students.

The remainder of the paper is structured as follows. First, we provide survey-relevant literature related to AI, LLM and ChatGPT. Following this, we present the theoretical foundation that grounds this paper and offer an overview of the methodological approach of how we gathered and analysed our data. In the subsequent section, we provide our findings and explore what they mean in terms of theoretical and practical contributions. We conclude by recognising the limitations of our study and suggesting potential avenues for future research.

## 2 Related Work

In recent years, AI has received considerable attention from both academia and practitioners with discussions from different vantage points (Berente et al. 2021; Enholm et al. 2021; Chatterjee et al. 2022; Grøder et al. 2022; Dwivedi et al, 2023; Haque et al. 2023). With the increase in LLM placing humans at the core of an AI ecosystem necessitates providing systematic guidance on AI development from developers or engineers, as well as interfaces to elucidate the consequences of their programming decisions (Zhou et al. 2023).

Due to the lack of research regarding using ChatGPT for accomplishing programming tasks among students, this narrative review section will summarise the extant literature and describe the current state of knowledge on the phenomenon. Particularly, the use of LLM in IS and CS education. This will be followed by a discussion on the application of ChatGPT as an intelligent system for completing programming exercises and assignments and will highlight the benefits and limitations of using ChatGPT in IS and CS education. This will be followed by the importance of developing critical programming skills among students and the potential risks associated with relying solely on ChatGPT for completing programming exercises and assignments. The insights that follow are derived from a multitude of studies, collated to provide a narrative review encapsulating previous observations on the subject matter (Schryen et al. 2020). It should be emphasised that the methodological rigour typically associated with a systematic literature review has not been adhered to in this instance.

### 2.1 Large Language Modules (LLM) in IS and CS Education

A notable characteristic of generative AI platforms is their rapid and widespread adoption; for instance, ChatGPT, launched on November 30, 2022, amassed a user base of one million within the initial five days post-release (Dwivedi et al. 2023; Zhou et al. 2023). The primary goal of ChatGPT is to generate natural language text for various applications (Sobania et al. 2023). The reciprocal relationship between humans and AI undamentally addresses the challenge of fostering trust in AI, particularly when it has not yet achieved full competence. By employing machines to enhance human cognitive abilities, humans have the potential to surpass their previous intellectual capacities. Additionally, AI can facilitate the expansion of human endeavours and render scarce human resources considerably more attainable. As an alternative, humans may delegate routine tasks to machines, relying on their efficiency in exchange for the opportunity to invest time and effort in more engaging, creative, or specialised activities. In both scenarios, machines can assist humans in amassing information and presenting various options while ultimately deferring to human users for the final decision-making (Dwivedi et al. 2023; Mikalef et al. 2023). Despite this, an overreliance on ChatGPT could introduce serious limitations to critical thinking and problem-solving, including but not limited to, IS and CS students. This will be discussed in the following section.

### 2.2 ChatGPT as an Intelligent System

A notable characteristic of generative AI platforms is their rapid and widespread adoption; for instance, ChatGPT, launched on November 30, 2022, amassed a user base of one million within the initial five days post-release (Dwivedi et al. 2023; Zhou et al. 2023). The primary goal of ChatGPT is to generate natural language text for various applications (Sobania et al. 2023). The reciprocal relationship between humans and AI fundamentally addresses the challenge of fostering trust in AI, particularly when it has not yet achieved full competence. By employing machines to enhance human cognitive abilities, humans have the potential to surpass their previous intellectual capacities. Additionally, AI can facilitate the expansion of human endeavors and render scarce human resources considerably more attainable. As an alternative, humans may delegate routine tasks to machines, relying on their efficiency in exchange for the opportunity to invest time and effort in more engaging, creative, or specialized activities. In both scenarios, machines can assist humans in amassing information and presenting various options while ultimately deferring to human users for the final decision-making (Dwivedi et al. 2023; Mikalef et al. 2023). Despite this, an overreliance on ChatGPT could introduce serious limitations to critical thinking and problem-solving, including but not limited to, IS and CS students. This will be discussed in the following section.

### 2.3 Limitations of ChatGPT

Though the internal mechanisms of ChatGPT remain unclear, OpenAI recognises certain limitations and researchers continue to engage with and scrutinize it to uncover further constraints (Else, 2023). Certainly, the impact on higher education has become a subject of debate, especially among IS and CS academics (Greene et al. 2022; Teubner et al. 2023, Dwivedi et al. 2023). Others argue that such tools can expedite knowledge acquisition, potentially benefiting the learning process. However, these platforms also present concerns regarding plagiarism and academic integrity, as students may employ them to complete assessments and dissertations (Stokel-Walker, 2022). Additionally, long term consequences may arise if students opt to forgo comprehensive and critical engagement with a topic, relying instead on ChatGPT for quick and potentially superficial learning and coding practices. More technical limitations will be provided in the experimental section.

### 2.4 Benefits of ChatGPT

As mentioned, ChatGPT is not limited to natural language (Dwivedi et al. 2023) and the tool is able to handle various querying languages such as Python, Java, C++ and others. For IS and CS programming students, OpenAI has now introduced various platforms that can suggest code and entire functions in real-time (e.g., GitHub Copilot, Codex). Despite this, it positively influences students and practitioners with low AI literacy or numerical skills (Zhou et al. 2023) as it provides the opportunity to invest time and effort in more engaging, creative, or specialized activities, especially for those students who do not major in these fields (e.g., Humanities). One major benefit of ChatGPT is the interaction with the system that allows for in-depth dialogue conversations to avoid misleading textual descriptions (Sobania et al. 2023). ChatGPT could be programmed to provide hints or suggestions that help programmers break down problems into more manageable parts. Additionally, ChatGPT could be designed to recognize common programming patterns and offer suggestions for how to apply them in new contexts, similar to how experts use their mental plans to solve problems. More technical benefits will be provided in the experimental section.

### 2.5 Development of Critical Programming Skills

Various academic institutions have called for more computational thinking to improve the transferability of programming skills in multiple courses (Matthee and Turpin, 2019). Despite Python being widely taught in both IS and CS courses, there is a notable lack of research and teaching guidelines in IS specifically focusing on problem-solving compared to CS. Others recommend that ChatGPT can be used in programming as a knowledge-creation tool for initial brainstorming (Megahed et al. 2023). The authors offered three distinct tasks encompassing (1) the comprehension of code, (2) the exploration of alternative methodologies for problem resolution through programming and (3) the translation of code between disparate programming languages (Megahed et al. 2023). The study further suggests that ChatGPT would alleviate the challenges among students associated with acquiring proficiency in programming languages, while simultaneously facilitating the sharing of code across various programming languages. A previous paper (De Raadt, Watson, and Toleman, 2006) advocates for the use of the Soloway (1986) model, to develop a tacit body of programming strategies to meet goals. For example, the model suggests that breaking down complex problems into smaller sub-problems is an effective problem-solving strategy.

### 2.6 Responsible Use of Generative AI (ChatGPT)

A paper by Dwivedi et al. (2023, p. 6) explores four themes under the umbrella of "Exploring the ethics of responsible AI: lessons from utilitarianism.": Theme 1: a tool view: the aim to improve the clarity of generative and self-adaptive AI models, to facilitate the explainability of resulting outputs (Haque et al. 2023); Theme 2: a proxy view: aims to understand ethical policies across organisations; Theme 3: a ensemble view: aims to understand under what conditions ChatGPT should be implemented; Theme 4: a skills view: aims to understand the skills and resources required to investigate the key limitations of ChatGPT across various contexts. Further studies suggested that the ongoing responsibility of academics is to foster critical thinking skills in students, a goal that has remained consistent over time. They also highlighted the potential for applications like ChatGPT to serve as companions or tutors in support of this objective (Dwivedi et al. 2023; Matthee and Turpin, 2019).

## 3 Theoretical approach

Complexity theory emerged within the domain of pure mathematics and was applied within the natural sciences to conceptualise better the non-linear and unpredictable nature of complex systems and the importance of feedback loops and emergence (McMillan, 2008; Stacey, 2003). More recently, the social sciences have increasingly adopted and adapted complexity theory to describe various social systems as complex systems (Moore et al., 2019; Brainard and Hunter, 2016; Hawe et al., 2009; Howarth et al., 2016; Moore and Evans, 2017; Rutter et al., 2017;). Within the field of education, complexity theory has been applied as a framework to understand the behaviour of complex systems such as the classroom environment (Jacobson et al. 2020).

Complex Adaptive Systems (CAS) theory is a branch of complexity theory that provides a novel framework for conceiving systems as constituted by autonomous agents that interact with each other following predetermined rules (Onik, Fielt & Gable, 2017). It facilitates an inquiry into the adaptive nature of complex systems and the emergence of order or properties that stem from the interactions of its constituent elements (Holland, 1995; Vidgen & Wang, 2006). The core components of a CAS include a) agents (individuals or entities that interact with each other), b) networks or connections between agents, and c) feedback loops or mechanisms that allow agents to learn and adjust their behaviour (Holland, 2006).

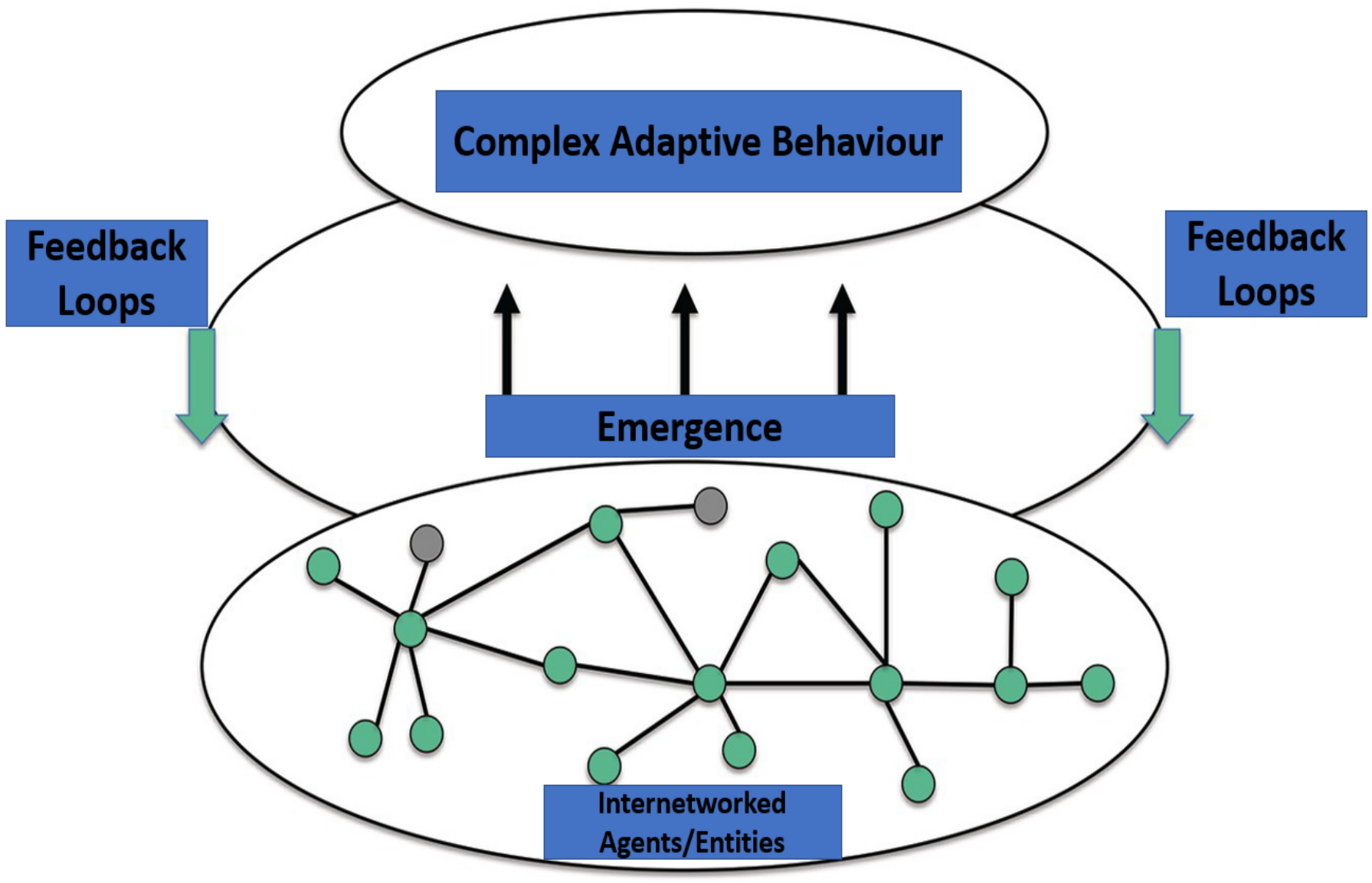

**Fig. 1.** Complex Adaptive System (adapted from van der Leeuw, 2020)

This research views the Information Systems programming classroom as a teaching and learning practice complex system i.e., students, lecturers/educators, programming tasks and the AI-driven technology (ChatGPT 3.5) as agents and entities interacting with each other within the classroom through the teaching and learning process of executing programming tasks. The interactions between individual entities (e.g., students, programming tasks, ChatGPT3.5) at the lower level create patterns in the classroom that emerge and are observable at the higher level (course outcomes), which, in turn, impact the interactions between individual entities. Onik, Fielt and Gable (2017) identified seven foundational constructs of the CAS theory, namely, 1) co-evolution, 2) emergence, 3) self-organisation, 4) fitness landscape, 5) the edge of chaos, 6) dynamism and non-linearity, 7) adaptation.

# 4 Methodology

The study adopted a pragmatist approach (Rorty, 2004) drawing on empirical evidence from real-world Python programming tasks from an introductory programming course aimed at first-year university students. Pragmatism emphasizes practicality, usefulness, and effectiveness and posits that ideas, theories, and beliefs should be evaluated based on their practical consequences and how they can be applied in real-world situations (Denzin, 2010; Rorty, 2004). Furthermore, pragmatism highlights the importance of experimentation, observation, and experience in evaluating the effectiveness of different approaches to solving problems. To ensure rigour, the paper followed the pragmatic qualitative analysis process suggested by Ramanadhan et al. (2021).

Firstly, a research question that addresses a practical problem was formulated, and then secondly, empirical evidence derived from our real-world situation was gathered. Thirdly, pragmatism suggests that ideas, theories, and beliefs should be evaluated based on their practical consequences. Fourthly, we applied our own experiences and the experiences of the participating students to gain insights into the implications of using ChatGPT in a programming classroom. Finally, pragmatism encourages an iterative approach to problem-solving. We conducted multiple iterations of different Python programming exercises to observe, analyse, explore, and refine our understanding of the implications of AI-led programming. Guided by the CAS constructs, we qualitatively and critically examine the dynamic and emergent interactions, pedagogical consequences and feedback loops due to the interaction of students of programming and ChatGPT (an AI-driven technology). We present several programming tasks that students are required to complete as part of learning new programming skills. We model how the freely available AI-driven application, ChatGPT 3.5, processes those tasks and discuss the consequent complex adaptive behaviours produced by this interaction.

.

# 5 Findings

In this section we requested ChatGPT to complete various programming tasks which are part of the Python exercises typically required for students to complete. The prompt and corresponding output from ChatGPT is displayed. ChatGPT has been asked to complete programming tasks for activity one based on questions in Fig.2. The prompt and the output from ChatGPT are presented as snapshots. The snapshots and responses represent the practical classroom interaction that students can make as part of completing their programming activities.

| No. | Name of Program | Output |
|---|---|---|
| 1 | escape_it.py | Python's syntax is nice. It's like writing in English. It is like they say it isn't 'hectic' |
| 2 | product.py | Enter the first number:<br>*2*<br>Enter the second number:<br>*4*<br>Product of 2 and 4 is: 8 |
| 3 | pass_fail.py | What is your course mark?<br>*50*<br>Passed |
| 4 | age.py | What is your age?<br>*20*<br>Adult |
| 5 | slicing.py | Enter your first name:<br>*Jane*<br>Enter your last name:<br>*Doe*<br>jan.doe |
| 6 | biggest.py | Enter the first number:<br>*1*<br>Enter the second number:<br>*3*<br>Enter the third number:<br>*2*<br>3 |

**Fig 2**. The six programming tasks for the first Python class activity.

In the next stage, see Fig. 3, the researchers prompted ChatGPT to complete the fourth programming activity and provided the output (Fig. 3). A Python file “my_bio.py” was created for this task by the chatbot.

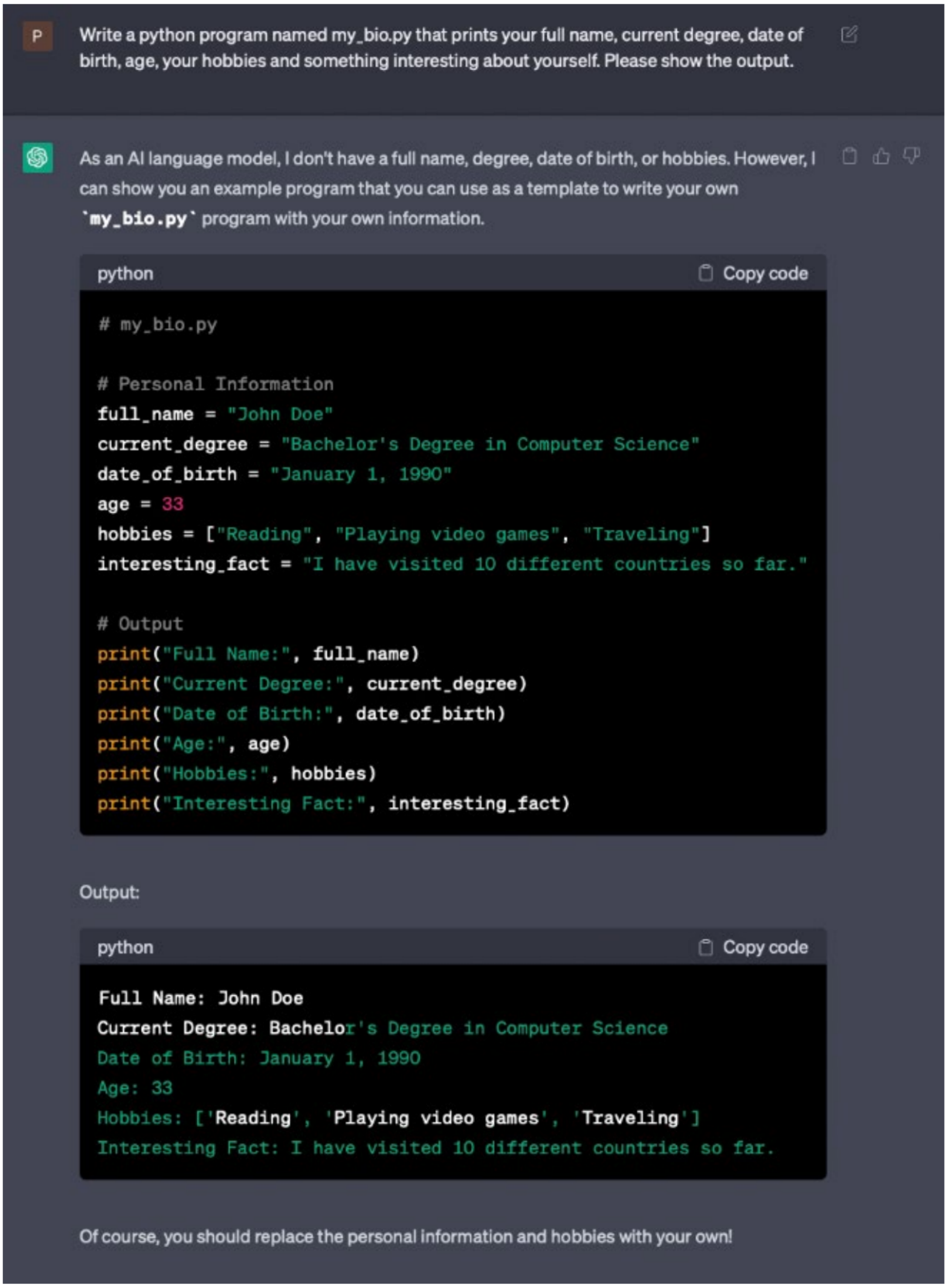

**Fig. 3**. ChatGPT output: prompt for the third Python task of activity one.

In Fig. 4, the researchers asked ChatGPT to complete the first programming activity by creating a “hello world” Python file and to print the output (on the right-hand side of Fig. 4). ChatGPT is prompted to complete the second programming activity, with an output displayed below in the form of an ASCII image (on the left-hand side of Fig. 4). The chatbot created the “ascii_art.py” Python file. ChatGPT presented a random smiley face as an output as it cannot read the web link provided to create the image. This presented a limitation.

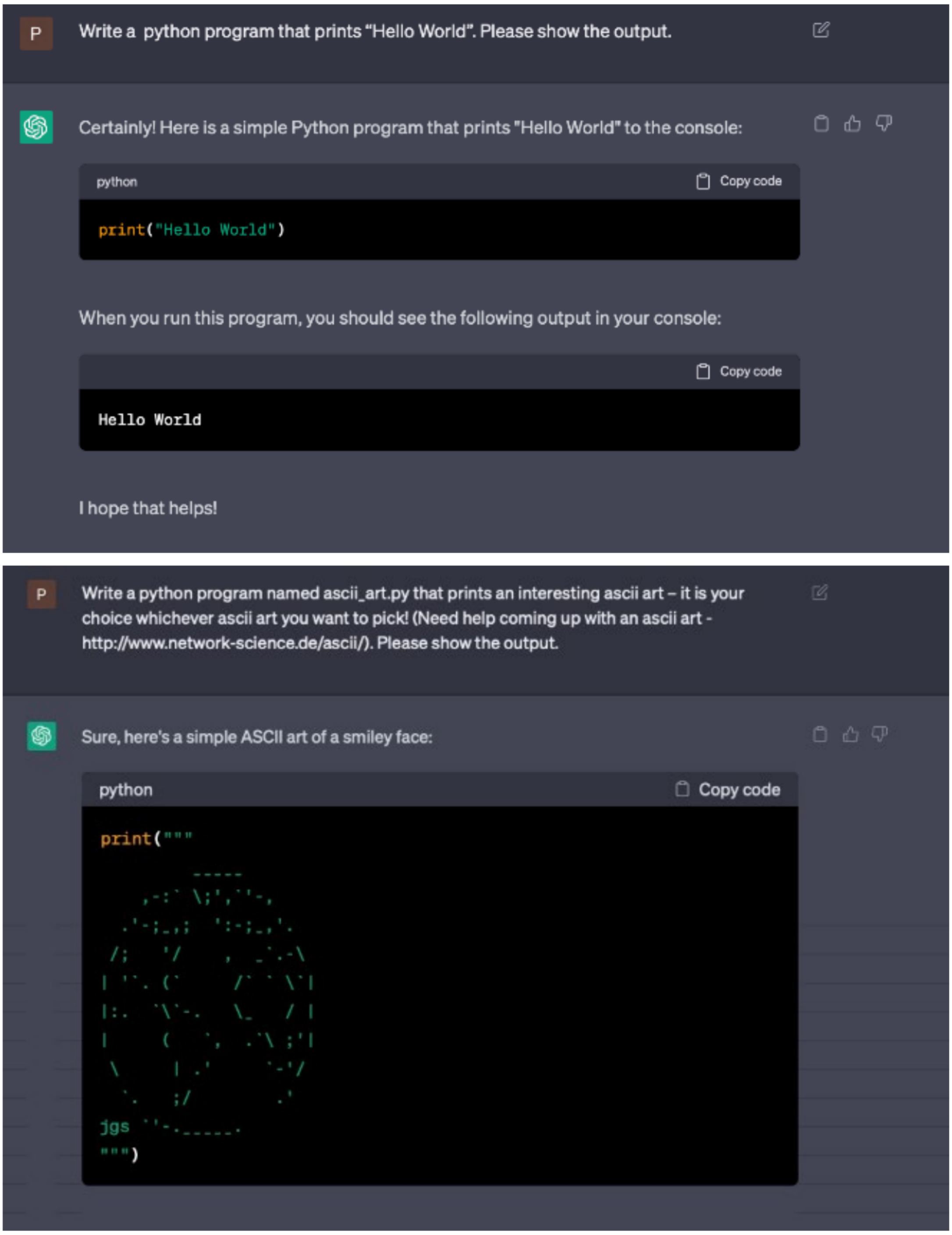

**Fig 4**. ChatGPT output: Create a file and print hello world (top) and create an ASCII image (bottom).

ChatGPT is then prompted to complete the fifth programming activity, the prompt and the output is displayed in Fig. 5. The chatbot created the "graduation_in.py" Python file and printed the output. The chatbot assumed that there are 365 days in a year and made this explicit in its output, this presents a possible limitation in its accuracy. ChatGPT is prompted to complete the sixth programming activity, the prompt and the output is displayed in Fig. 6. The chatbot created the "flight_time.py" Python file and printed the output.

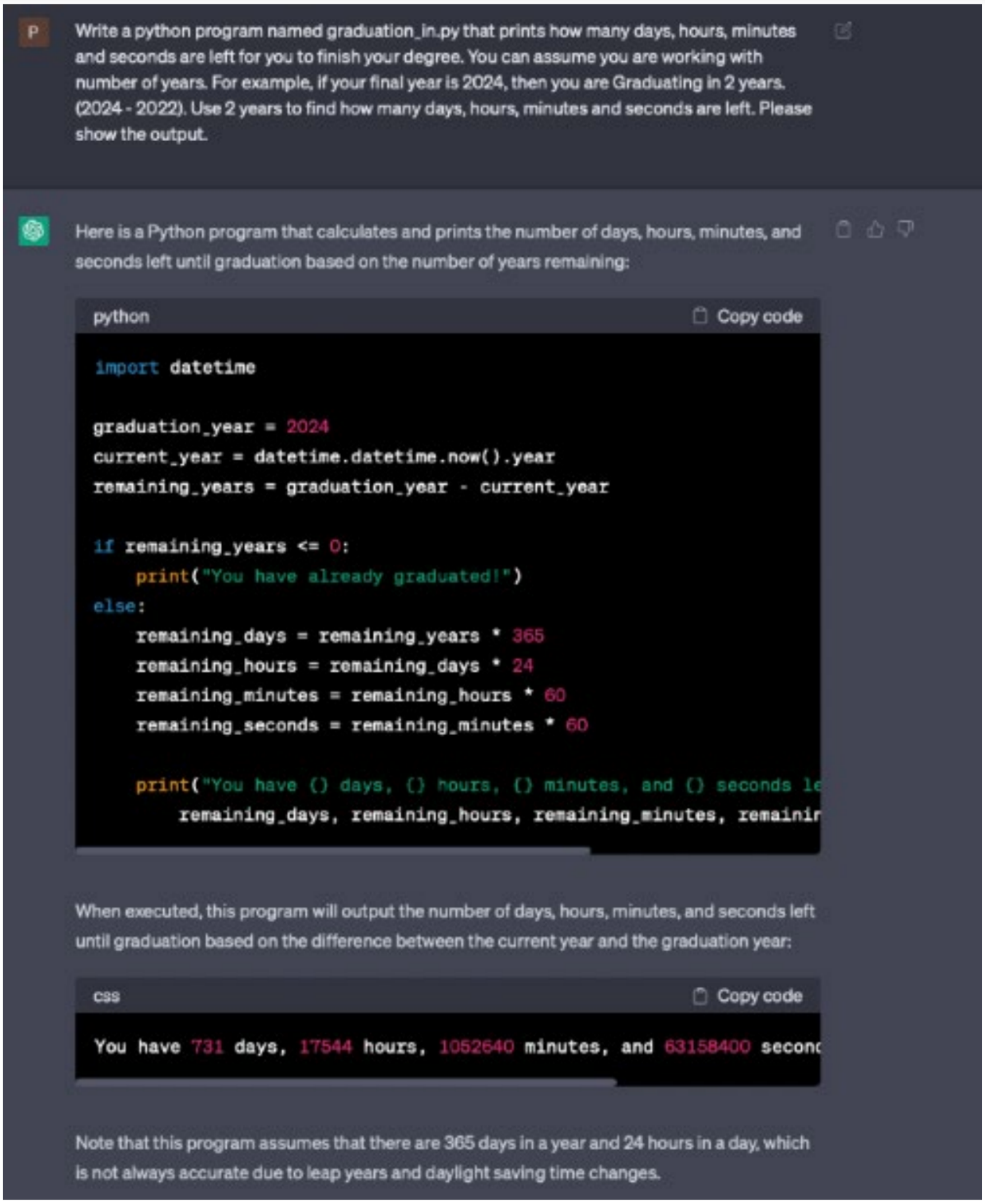

**Fig. 5.** ChatGPT output: prompt for the fifth Python task of activity one.

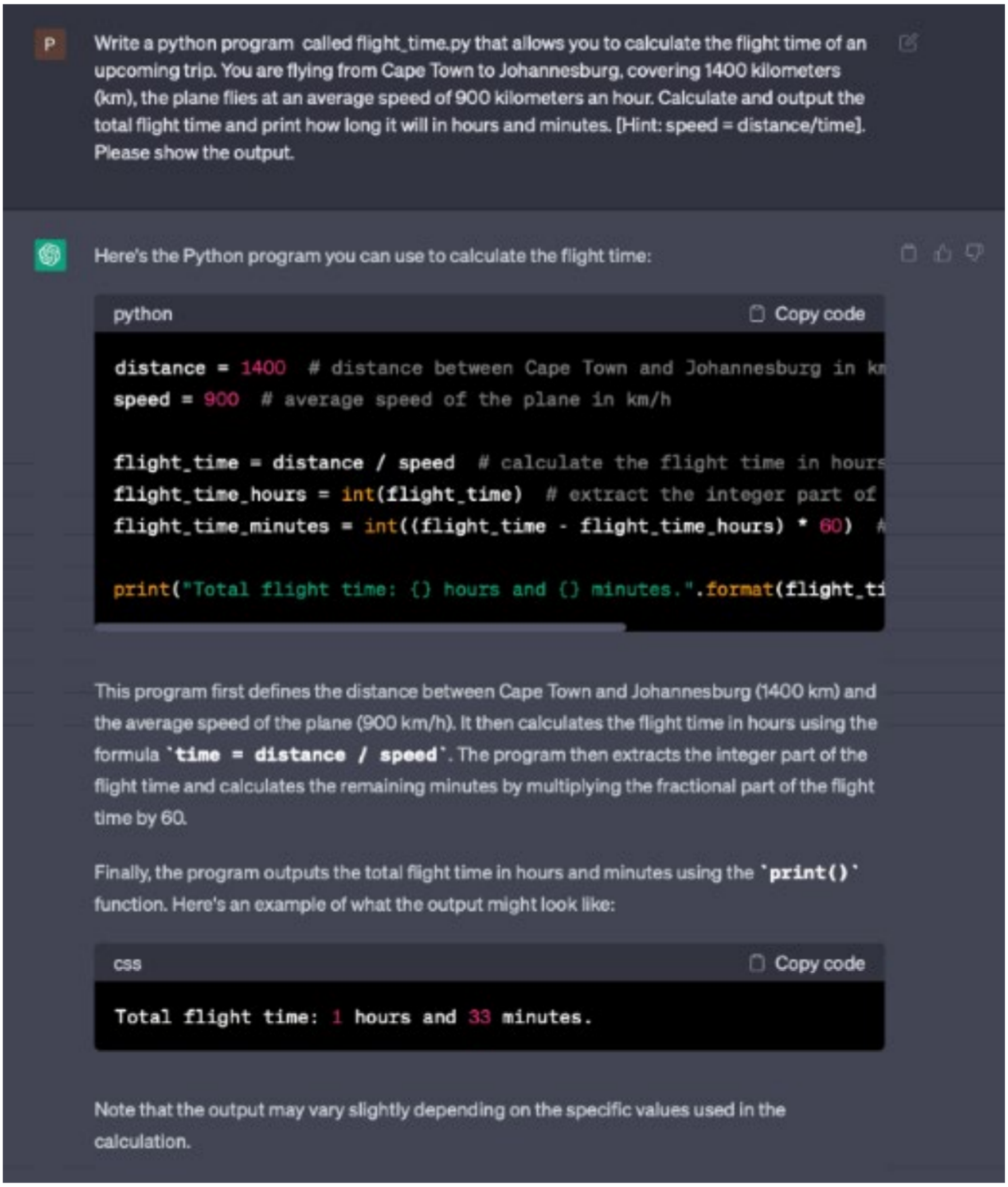

**Fig. 6.** ChatGPT output: prompt for the sixth Python task of activity one.

In Fig. 7, the expected output for all the tasks for the first programming activity are shown, the prompted task closely resembled the expected outcome for each of these tasks.

| No. | Name of Program | Output |
|---|---|---|
| 1 | hello_world.py | Hello World |
| 2 | my_bio.py | Name: John Doe<br>Current Degree: Bachelors in Commerce Specializing in Information Systems<br>Date of Birth: 31 December 2000<br>Age: 22<br>Hobbies: Espresso and Chill<br>Something interesting about myself: I hate questions such as these! |
| 3 | ascii_art.py | 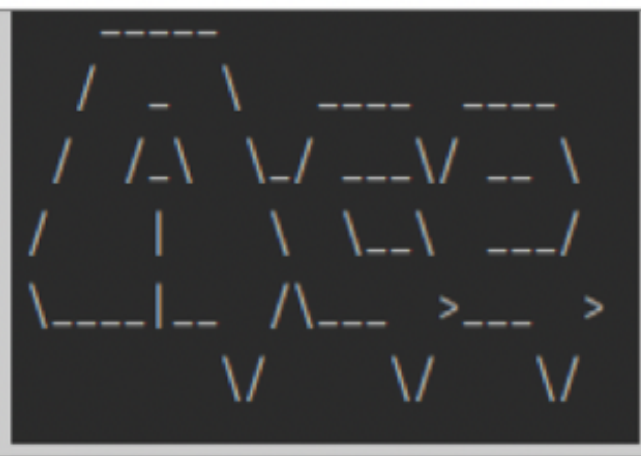 |
| 4 | my_bio_2.py | Name: John Doe<br>Current Degree: Bachelors in Commerce Specializing in Information Systems<br>Date of Birth: 31 December 2000<br>Age: 22<br>Hobbies: Espresso and Chill<br>Something interesting about myself: I hate questions such as these! |
| 5 | graduation_in.py | Days Left: 730<br>Hours Left: 17520<br>Minutes Left: 1051200<br>Seconds Left: 63072000 |
| 6 | flight_time.py | It will take: 1.0 hour(s) and 33.333333333333336 minute(s) to arrive. |

**Fig 7**. The expected course output for Python activity one.

ChatGPT has furthermore been asked to complete programming tasks for activity two based on questions in Fig. 8. The prompts and outputs are displayed below.

| No. | Name of Program | Question |
|---|---|---|
| 1 | escape_it.py | Write a program that prints "Python's syntax is nice. It's like writing in English. It is like they say it isn't 'hectic'". |
| 2 | product.py | Write a program that asks the user to enter 2 numbers. The program then prints the sum and product of the 2 numbers. |
| 3 | pass_fail.py | Write a program that asks the user to enter their course mark. Print "Passed" if the mark is 50 or higher, otherwise print "Failed" |
| 4 | age.py | Write a program that asks the user to enter their age. Print "baby" if the age is between 0 and 4 inclusive, print "child" if the age is between 5 and 12 inclusive, print "teenager" if the age is between 13 and 19 inclusive and print "adult" if the age is 20 or older. |
| 5 | slicing.py | Write a program that asks the user to enter the first name and then their last name. Print the first 3 characters of the first name, a dot (".") and the last 3 characters of the last name. All of the characters need to be lower-case when you print them.<br><br>E.g., If the name is John Doe, you will print: joh.doe |
| 6 | biggest.py | Write a program that asks the user to enter 3 numbers, print the biggest number. |

**Fig 8**. Instructions for tasks of the second Python activity.

ChatGPT is prompted to complete the first activity of the second programming activity, the prompt and the output is displayed in Fig. 9. The chatbot created the "escape_it.py" Python file and printed the output.

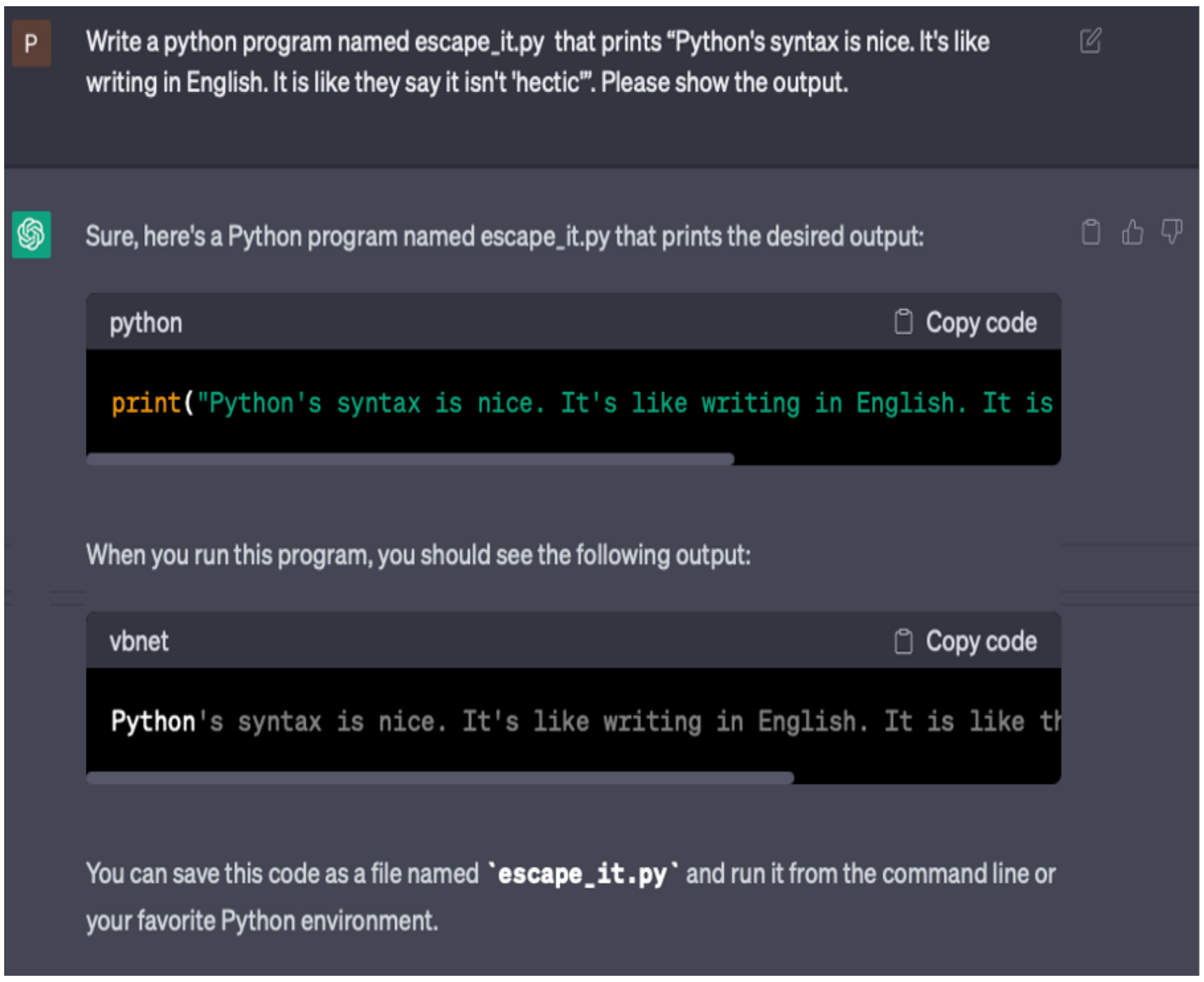

**Fig 9**. ChatGPT output: prompt for the first Python task of the second activity.

In the snapshot in Fig. 10, ChatGPT is prompted to complete the second task of programming activity two and provides the output. A Python file for this task was created by the chatbot. ChatGPT also provides context on the functions used in the coding of the program, this provides benefits for students learning the specific programming language, this is an example of a positive feedback loop.

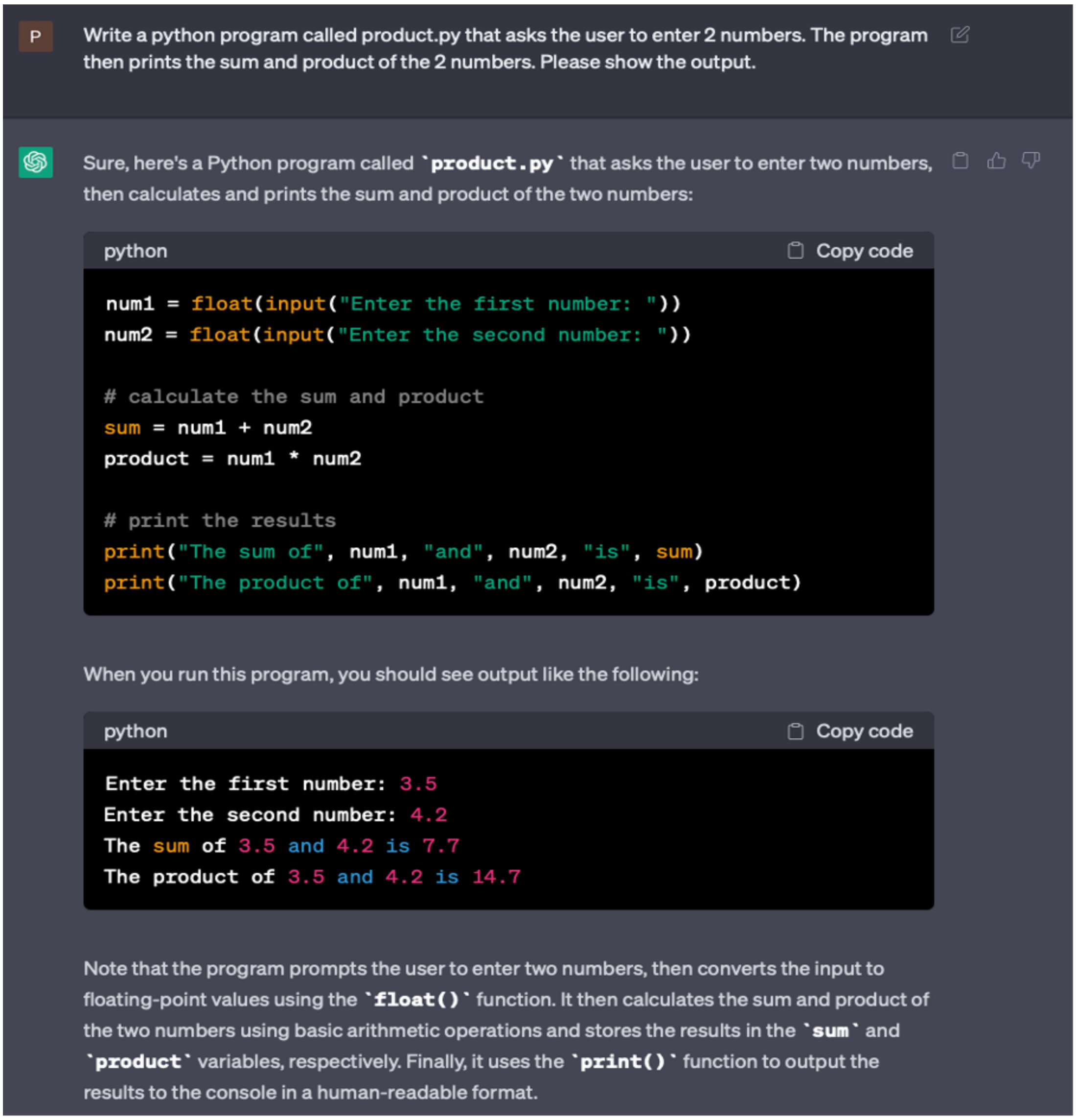

**Fig 10**. ChatGPT output: prompt for the second Python task of the second activity.

Fig. 11 shows a snapshot of the ChatGPT prompt to complete the third task of programming activity two and provides the output. The chatbot created a Python file for this task.

In the snapshot in Fig. 12, ChatGPT is prompted to complete the fourth task of programming activity two and provides the output. The chatbot created a Python file for this task. This output is another demonstration of ChatGPT providing positive feedback as it shows students how it utilised functions in its output.

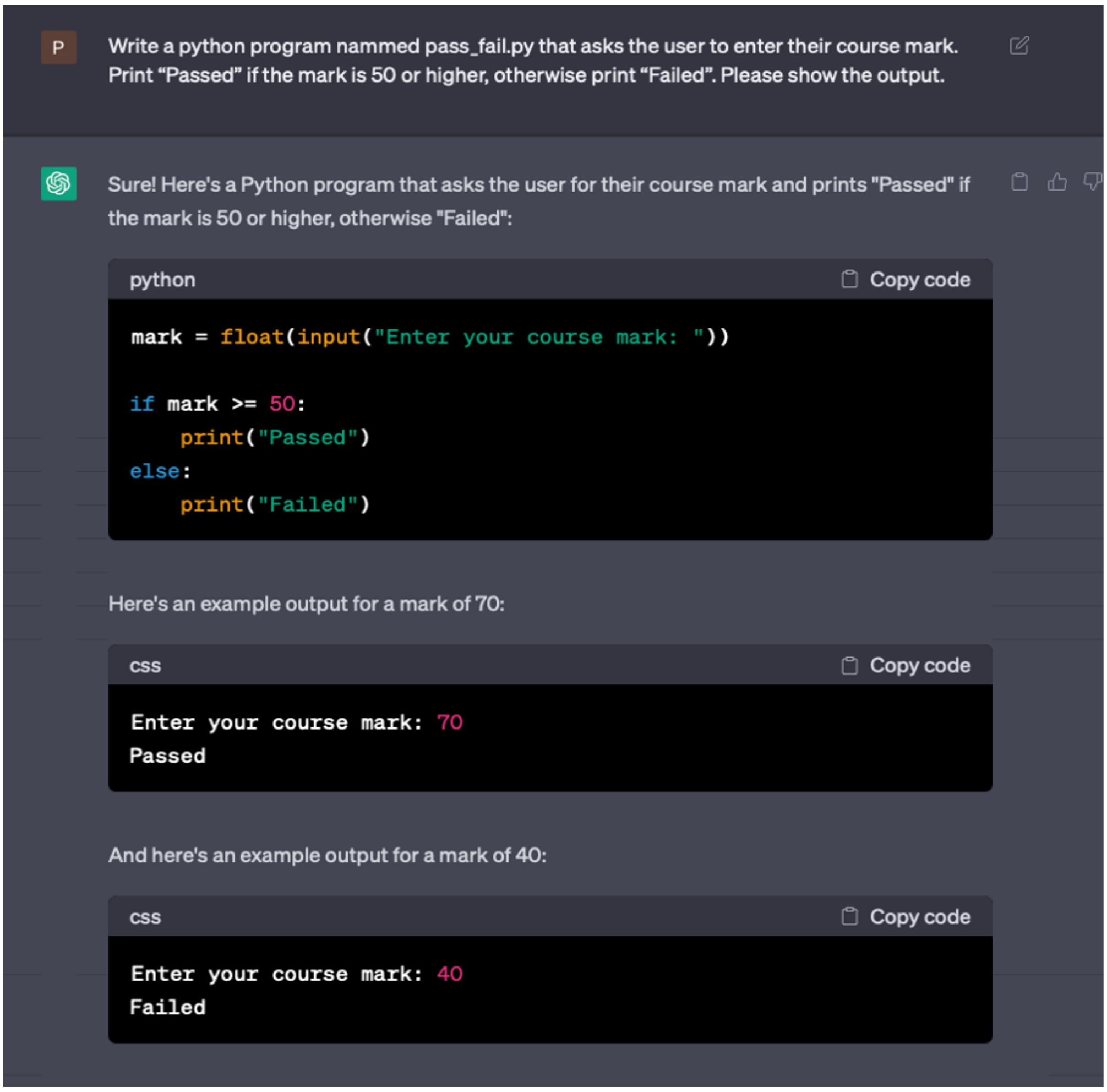

**Fig 11**. ChatGPT output: Prompt for the third Python task of the second activity.

In the snapshot in Fig. 12, ChatGPT is prompted to complete the fourth task of programming activity two and provides the output. A Python file for this task was created by the chatbot. This output is another demonstration of ChatGPT providing positive feedback as it shows students how it utilised functions in its output.

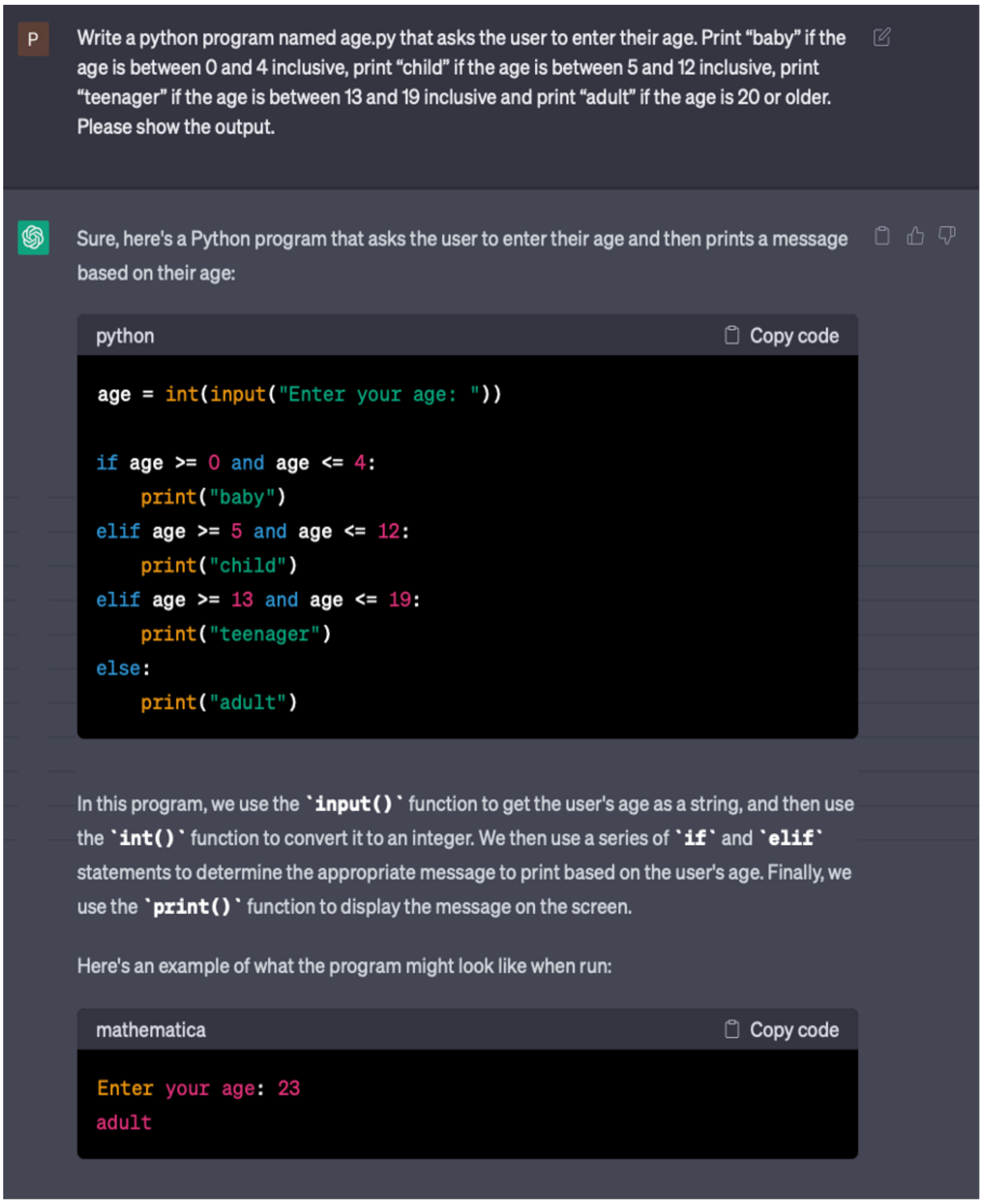

**Fig 12**. ChatGPT output: prompt for fourth Python instruction as part of the second activity.

The snapshot in Fig. 13 displays the prompt in ChatGPT of the second task of programming activity two and provides the output with positive feedback, demonstrating more benefits of ChatGPT to students learning this programming language.

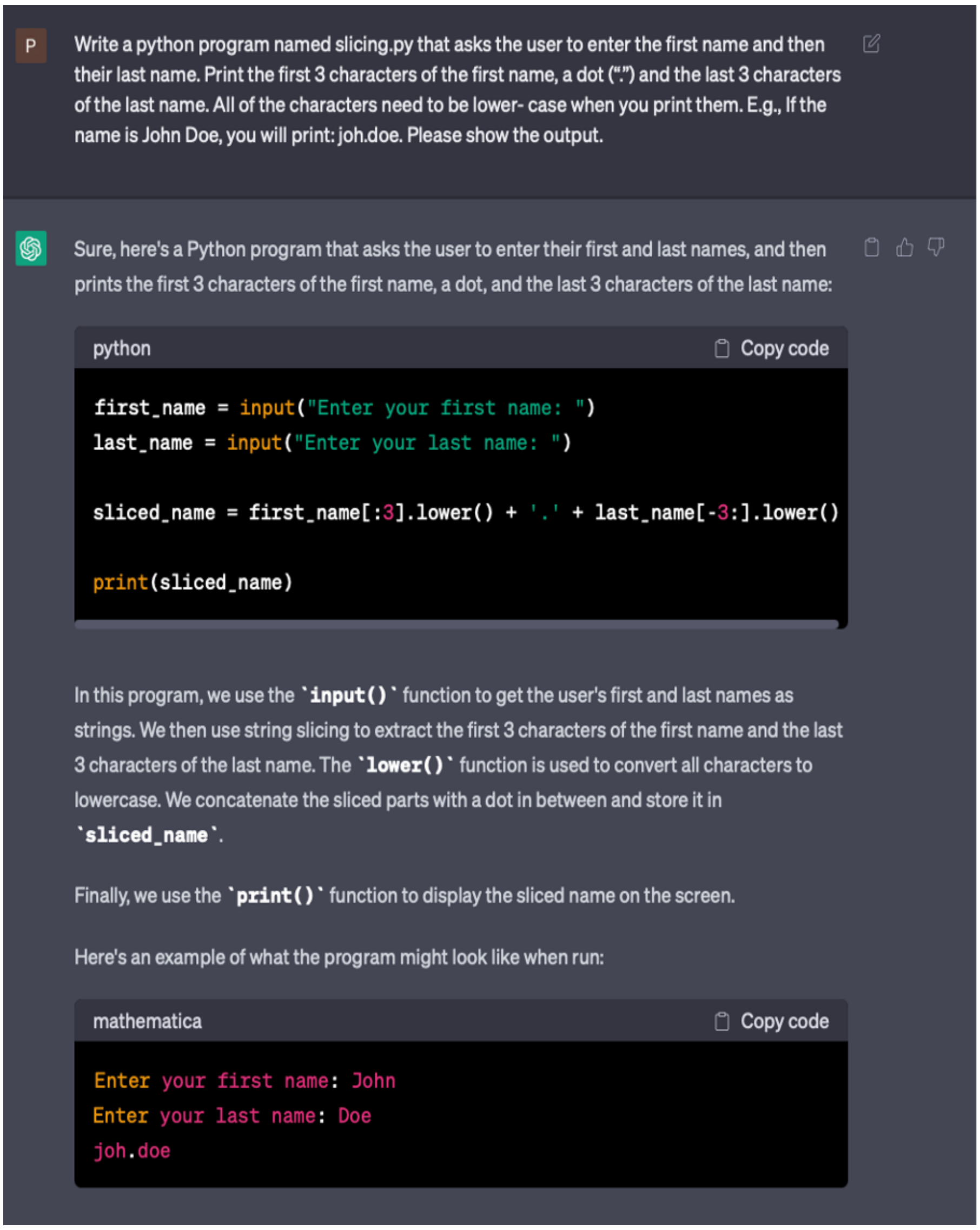

**Fig 13**. ChatGPT output: Prompt for the fifth Python task of the second activity.

In the snapshot in Fig. 14, ChatGPT is prompted to complete the sixth task of programming activity two and provides the output with positive feedback, further demonstrating its benefits to students.

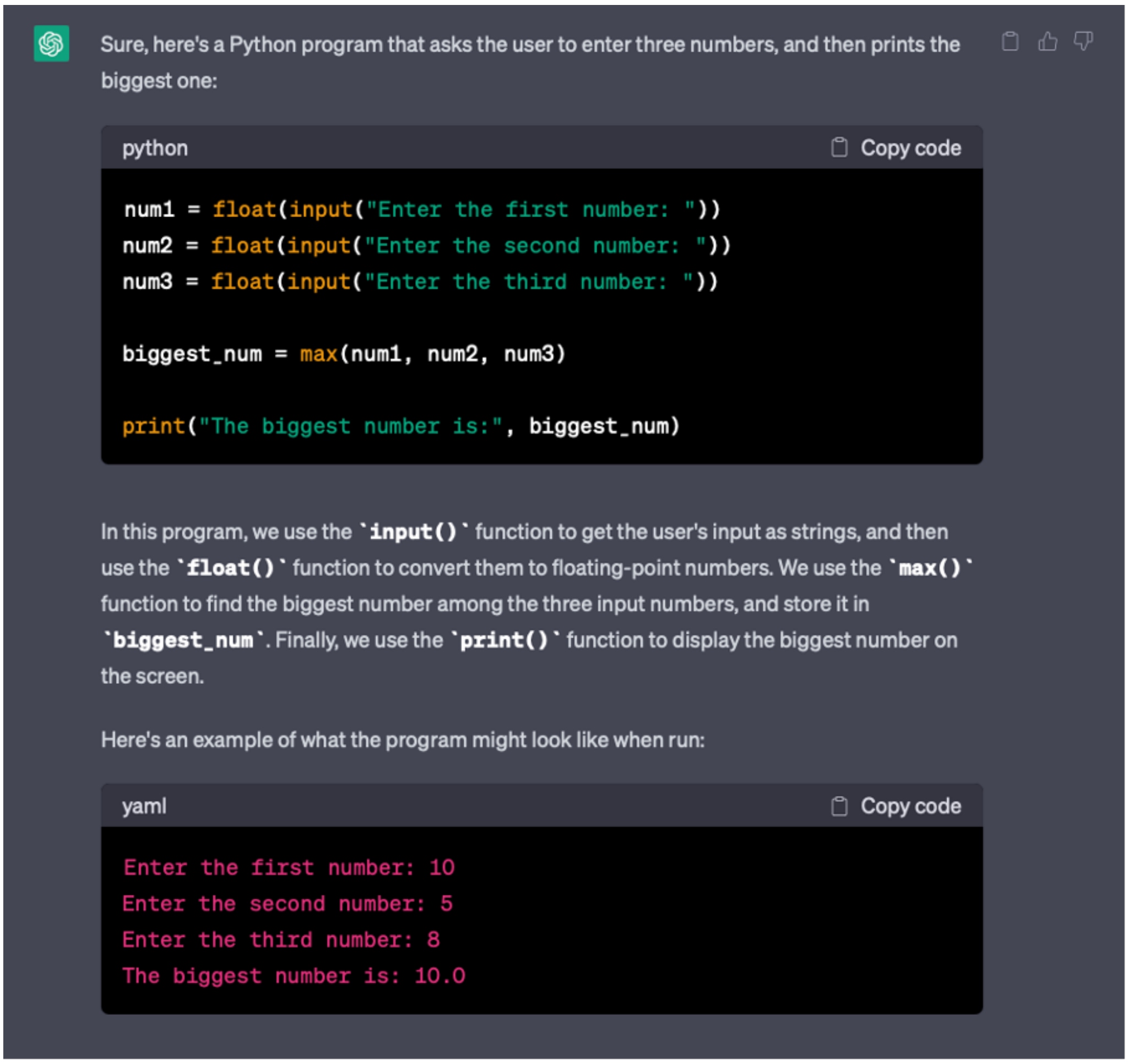

Sure, here's a Python program that asks the user to enter three numbers, and then prints the biggest one:

python Copy code

```python
num1 = float(input("Enter the first number: "))
num2 = float(input("Enter the second number: "))
num3 = float(input("Enter the third number: "))

biggest_num = max(num1, num2, num3)

print("The biggest number is:", biggest_num)
```

In this program, we use the `input()` function to get the user's input as strings, and then use the `float()` function to convert them to floating-point numbers. We use the `max()` function to find the biggest number among the three input numbers, and store it in `biggest_num`. Finally, we use the `print()` function to display the biggest number on the screen.

Here's an example of what the program might look like when run:

yaml Copy code

```yaml
Enter the first number: 10
Enter the second number: 5
Enter the third number: 8
The biggest number is: 10.0
```

**Fig 14**. ChatGPT prompt for the sixth Python task of second activity with the output.

In Fig. 15 the expected exercise output is shown, the prompted tasks closely resembled the expected outcome for each of these tasks.

| No. | Name of Program | Output |
|---|---|---|
| 1 | escape_it.py | Python's syntax is nice. It's like writing in English. It is like they say it isn't 'hectic' |
| 2 | product.py | Enter the first number:<br>*2*<br>Enter the second number:<br>*4*<br>Product of 2 and 4 is: 8 |
| 3 | pass_fail.py | What is your course mark?<br>*50*<br>Passed |
| 4 | age.py | What is your age?<br>*20*<br>Adult |
| 5 | slicing.py | Enter your first name:<br>*Jane*<br>Enter your last name:<br>*Doe*<br>jan.doe |
| 6 | biggest.py | Enter the first number:<br>*1*<br>Enter the second number:<br>*3*<br>Enter the third number:<br>*2*<br>3 |

**Fig 15**. Python exercise output: Activity 2.

# 6 Discussion

A key observation emanating from a typical interaction between the student and ChatGPT is that the AI-driven application provides perfect solutions to the problems presented through merely copying and pasting from the activity sheet. The technology agent therefore takes over the expected learning, effort and experience that the student is expected to undergo. This observation presents an emergent and existential conundrum: If the machine is learning, is the student learning?

In CAS, co-evolution refers to the dynamic process where two or more agents evolve together in response to each other's actions. The co-evolutionary process in a classroom can lead to the development of complex adaptations that allow students to interact with AI technology (ChatGPT) in more complex and sophisticated ways. In our case, while the machine (Agent B) evolves in its better understanding of answering the student, there is a likelihood that the student is developing dependency and disempowerment as they lose out on the opportunity to develop practical programming skills. Emergence in CAS refers to the phenomenon where complex patterns, structures, or behaviours spontaneously arise from the interactions between many simple agents within the system. Emergence in this study refers to new behaviours such as dependency (over-reliance) on the AI tool and loss of competency development that can arise when students rely on technology instead of internalising the learning.

Self-organisation in CAS is the potential to spontaneously form patterns or structures without being directed by any external force or central authority. This emergent behaviour arises from the interactions between the system's individual components. In the interaction with the student and ChaptGPT, while the AI might tend towards self-organising during its interaction with Agent A - the student, it is not evident that the student will grow from this interaction. Due to the pressure and demands of university studies, when students obtain answers for the work assigned to them, they are likely to consider the work to be complete and simply move on to the next urgent tasks. Tasks such as the ones shown in the previous section, are aimed at slowing the student down to spend time engaging with the material. However, this opportunity is lost when they have a powerful answer generator to rely on.

The edge of chaos in CAS, applied to our context, is the state where the teaching and learning practice is balanced between order (undisrupted teaching and learning) and chaos (disrupted teaching and learning). In this state of order (the edge of chaos), the system has enough order to be stable and predictable but enough chaos to allow for flexibility and adaptation to changing conditions. We propose that the introduction of the ChatGPT agent generates a bias towards chaos as the student is deprived of achieving and experiencing the intended learning outcomes. Dynamism in CAS refers to the system's ability to respond to the environment in real-time and to continuously adjust behaviour and structure in response to changing conditions. Non-linearity specifies that small changes in one part of a CAS can lead to significant, unpredictable effects throughout the system. Together, dynamism and non-linearity posit the classroom interaction between the students and the AI (ChatGPT) as highly complex and challenging to predict or control. We find that educators (Agent C) will experience high levels of difficulty in monitoring student progress. This presents a significant challenge to the teaching and learning process. The dynamic and non-linear interactions between the three primary agents leads to the emergence of unintended negative complex adaptive behaviors overall.

# 7 Conclusion and Future Work

As ChatGPT becomes more widely utilised for activities like programming there will be a heightened emphasis on accountability when teaching and learning issues arise. The increasing reliance on ChatGPT has the potential to lead to a shift in educational expectations and outcomes, as pressure to rethink traditional teaching methods and pedagogies develops. This will have profound implications for common assessment practices, marking guidelines and academic recommendations for IS and CS education. From the examples this paper covered, while the introduction of ChatGPT has the potential to provide benefits such as positive feedback loops (providing context to programming solutions), it also generates concerning negative feedback loops of over-reliance and disempowerment. With OpenAI's API being widely used for code generation, students can now produce code without any deep computational knowledge, creating a new wave of ChatGPT-assisted programmers and developers (“prompt engineers”) which might lead to unintended consequences such as over-reliance on AI and the decline of computational skills development. This raises questions about the impact on performative abilities and the value of entering exam locations without prior knowledge, as ChatGPT is mainly used for assignments and tutorials. Consequently, this trend may affect student results and lead to potential demand for "master intelligent prompters" as a new industry skill. However, this reliance on ChatGPT could erode essential IS and CS skills, as ChatGPT may generate incorrect code or negatively impact programming projects. This research contributes a cautious reflection to the emergent discourse regarding the use of LLMs in the classrooms, in particular, the programming classrooms. Future work for this work should involve longitudinal observational studies that would compare, through mixed method inquiry, how the computational thinking skills of students develop over time as various AI-driven programming tools become available to students.